\documentclass{article}

\usepackage[preprint, nonatbib]{nips_2018}

\usepackage{float}

\usepackage[square,numbers]{natbib}
\usepackage[utf8]{inputenc} 
\usepackage[T1]{fontenc}    
\usepackage{url}            
\usepackage{booktabs}       
\usepackage{amsfonts}       
\usepackage{nicefrac}       
\usepackage{microtype}      
\usepackage{graphicx}
\usepackage{tabularx}

\usepackage{amsmath}
\usepackage{algorithm}  
\usepackage{algpseudocode} 
\usepackage{subfigure}
\usepackage{multirow}

\title{HCAF-DTA: drug-target binding affinity prediction with cross-attention fused hypergraph neural networks}

\author{
	Jiannuo Li \\
        School of Mathematics, University of Hunan \\
        Changsha 410082, China \\
	\texttt{jiannuo@hnu.edu.cn} \\
	\And
	Lan Yao \\
        School of Mathematics, University of Hunan \\
        Changsha 410082, China \\
	\texttt{yao@hnu.edu.cn} \\
}

\begin{document}
\maketitle

\begin{abstract}
	
Accurate prediction of the binding affinity between drugs and target proteins is a core task in computer-aided drug design. Existing deep learning methods tend to ignore the information of internal sub-structural features of drug molecules and drug-target interactions, resulting in limited prediction performance. In this paper, we propose a drug-target association prediction model HCAF-DTA based on cross-attention fusion hypergraph neural network. The model innovatively introduces hypergraph representation in the feature extraction stage: drug molecule hypergraphs are constructed based on the tree decomposition algorithm, and the sub-structural and global features extracted by fusing the hypergraph neural network with the graphical neural network through hopping connections, in which the hyper edges can efficiently characterise the functional functional groups and other key chemical features; for the protein feature extraction, a weighted graph is constructed based on the residues predicted by the ESM model contact maps to construct weighted graphs, and multilayer graph neural networks were used to capture spatial dependencies. In the prediction stage, a bidirectional multi-head cross-attention mechanism is designed to model intermolecular interactions from the dual viewpoints of atoms and amino acids, and cross-modal features with correlated information are fused by attention. Experiments on benchmark datasets such as Davis and KIBA show that HCAF-DTA outperforms state of the arts in all three performance evaluation metrics, with the MSE metrics reaching 0.198 and 0.122, respectively, with an improvement of up to 4\% from the optimal baseline.
	
\end{abstract}

\section{Introduction}

Drug discovery is a complex and time-consuming systematic engineering endeavor. Statistical data reveal that developing a successful new drug requires an average investment of approximately \$2.6 billion and spans a 13.5-year research and development cycle\cite{2010How}. Against this backdrop, scientists urgently need to explore more efficient strategies for drug discovery. Drug-target interaction (DTI) lies at the core of this process, as its study is not only critical for novel drug development but also holds profound implications for predicting and understanding potential drug side effects.

In recent years, computer-aided drug design (CADD) has emerged as a pivotal tool in drug discovery due to its cost-effectiveness and high efficiency, garnering increasing attention. Early studies predominantly employed traditional machine learning methods for drug-target affinity (DTA) prediction. For instance, KronRLS\cite{pahikkala2015toward} calculated drug-drug and protein-protein relationships via Kronecker product operations to generate a kernel matrix, followed by least squares regression (RLS). SimBoost\cite{he2017simboost} defined three types of features for drugs, proteins, and drug-protein pairs, respectively, and utilized boosting algorithms for prediction.

Despite their success, these methods heavily rely on labor-intensive feature engineering. To address this limitation, deep learning approaches have been increasingly adopted for DTA prediction. The core advantage of deep learning lies in its ability to automatically extract critical features from massive datasets, circumventing the need for manual feature engineering. In DTA tasks, models typically process drug SMILES sequences and protein amino acid sequences as inputs. For example, DeepDTA\cite{ozturk2018deepdta} innovatively applied convolutional neural networks (CNNs) to encode drug SMILES and protein sequences, feeding the encoded features into fully connected networks for prediction. WideDTA\cite{ozturk2019widedta} extended this framework by integrating protein interaction domains and ligand maximum common substructures, employing four CNN modules to enhance performance. However, CNNs alone struggle to comprehensively capture sequential dependencies. To dynamically model atomic relationships and assign weights, attention mechanisms were introduced. AttentionDTA\cite{zhao2019attentiondta} incorporated attention layers atop CNN encoders to focus on key sequence regions, further improving accuracy. DrugBAN\cite{bai2023interpretable} proposed a bilinear attention network to weight feature interactions, while HyperAttentionDTI\cite{zhao2022hyperattentiondti} combined CNNs with attention vectors for atom- and residue-level representations. FusionDTA\cite{yuan2022fusiondta} replaced traditional pooling with multi-head linear attention to aggregate global features.

Notably, SMILES and protein sequences inherently lack spatial information, which is crucial for understanding molecular design and protein conformation. To address this, graph neural networks (GNNs) have been leveraged to capture spatial dependencies. GraphDTA\cite{nguyen2021graphdta} encoded drug SMILES into undirected graphs using RDKit and processed them via GNNs, achieving significant performance gains. DGraphDTA\cite{jiang2020drug} introduced weighted protein graphs to model drug-protein pairs, while WGNN-DTA\cite{jiang2022sequence} utilized ESM-1b-predicted residue contact maps to construct edge-weighted protein graphs. TrimNet-CNN\cite{li2021trimnet} employed triplet message passing for molecular representation learning, and EmbedDTI\cite{jin2021embeddti} integrated atom and substructure graphs with attention-augmented graph convolutional networks. MGraphDTA\cite{yang2022mgraphdta} designed a 27-layer deep GNN with skip connections to mitigate gradient vanishing and over-smoothing. GLCN-DTA\cite{qi2024drug} incorporated graph learning-convolutional networks, and GSAML-DTA\cite{liao2022gsaml} fused GAT-GCN architectures with mutual information optimization. AMMVF-DTI\cite{wang2023ammvf} explored node- and graph-level embeddings via multi-head self-attention, whereas MFR-DTA\cite{hua2023mfr} combined BioMLP/CNN and GNN/LSTM modules with cross-attention.

Despite these advancements, existing DTA prediction methods often simply concatenate drug and protein features for final predictions, neglecting intricate intermolecular interactions and limiting generalization in cold-start scenarios. To overcome these limitations, we propose HCAF-DTA, a hypergraph neural network model with bidirectional cross-attention fusion. In the feature extraction phase, HCAF-DTA constructs drug hypergraphs via tree decomposition, integrating hypergraph and simple graph representations through skip connections to capture both substructural and global molecular features. For proteins, residue contact maps predicted by ESM-1b are converted into weighted graphs, and multi-layer GNNs extract spatial dependencies. During prediction, a bidirectional multi-head cross-attention mechanism simulates atom-residue interactions from dual perspectives, dynamically fusing features to model binding affinity. This approach preserves critical interaction patterns, enhancing robustness in cold-start settings.
\section{Method}
\begin{figure}[htbp]
    \centering    \includegraphics[width=1\linewidth]{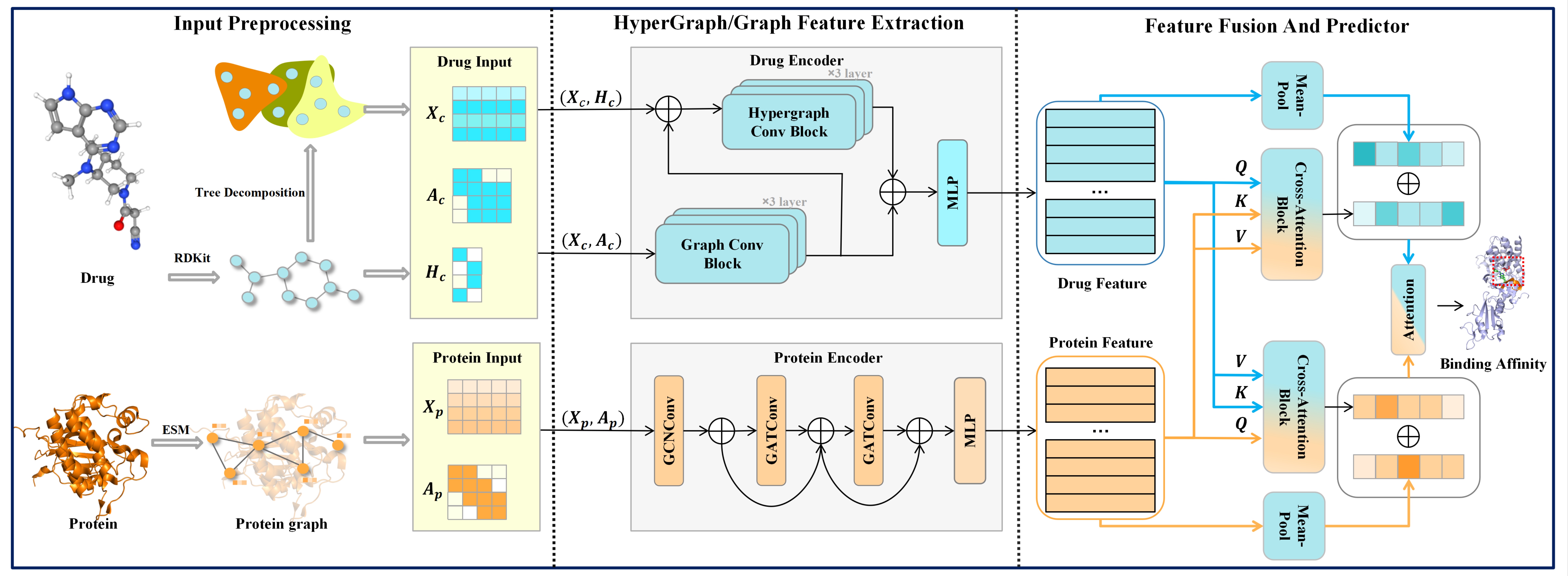}
    \caption{The architecture of our proposed HCAF-DTA mode}
    \label{fig:model2}
\end{figure}

This paper proposes HCAF-DTA, an end-to-end prediction algorithm for drug-target affinity (DTA) prediction. The overall architecture of HCAF-DTA is illustrated in Figure \ref{fig:model2}, comprising three core components: (1) a data preprocessing module, (2) drug and protein encoders for feature extraction (HyperGraph/Graph Feature Extraction), and (3) a feature fusion module with bidirectional multi-head cross-attention for affinity prediction (Feature Fusion and Predictor).

Firstly, the model processes drug SMILES sequences and protein FASTA sequences to generate molecular representations, including simple graphs and hypergraphs for drugs, along with graph representations for proteins. Subsequently, dedicated drug and protein encoders extract sophisticated spatial information features to obtain comprehensive representations. The model then employs a dual-perspective feature fusion mechanism to capture interaction patterns between individual atoms and amino acids, ultimately integrating these interactions through attention mechanisms for final prediction via MLP layers.

For a given drug SMILES sequence, preprocessing yields three components: a feature matrix $X_c \in \mathbb{R}^{n_c \times d_c}$, an adjacency matrix $A_c \in \mathbb{R}^{n_c \times n_c}$, and a hyperedge matrix $H_c \in \mathbb{R}^{n_c \times m_h}$, where $n_c$ denotes atom count, $d_c$ represents feature dimension, and $m_h$ indicates hyperedge count. For protein FASTA sequences, preprocessing generates a feature matrix $X_p \in \mathbb{R}^{n_p \times d_p}$ and an adjacency matrix $A_p \in \mathbb{R}^{n_p \times n_p}$, with $n_p$ representing amino acid count and $d_p$ denoting feature dimension. As shown in Figure \ref{fig:model2}, the drug encoder $f_c(\cdot)$, protein encoder $f_p(\cdot)$, and fusion-prediction module $f_{\text{cross}}(\cdot)$ operate as follows:
\begin{align}
    Z_c &= f_c(X_c,A_c,H_c),\nonumber\\
    Z_p &= f_p(X_p,A_p),\nonumber \\
    \tilde{y} &= f_{cross}(Z_c,Z_p).\nonumber
\end{align}
\subsection{Substructural Feature Extraction in Hypergraph Perspective}
Traditional graph-based methods effectively represent pairwise interactions between nodes in simple graphs. However, nodes often exhibit complex non-pairwise relationships that go beyond pairwise interactions. Representing these complex higher-order relationships using only graph-based methods may lead to information loss or redundancy. For example, in drug molecules, there exist complex substructures and functional groups that represent higher-order relationships. To address this issue, the model introduces hypergraph structures. Hyperedges in hypergraphs can contain multiple nodes, enabling hypergraphs to encode many-to-many interactions among multiple nodes. Hypergraph neural networks update node representations by aggregating information from nodes within hyperedges through hyperedge convolution layers, thereby capturing complex relationships between nodes. This aggregation process ensures that the features of each node not only include its own information but also integrate information from other nodes connected via hyperedges. Consequently, through feature propagation within hyperedges and information exchange between hyperedges, hypergraphs naturally represent the complex relationships between substructures in drug molecules.

\begin{figure}[htbp]
    \centering
    \includegraphics[width=1\linewidth]{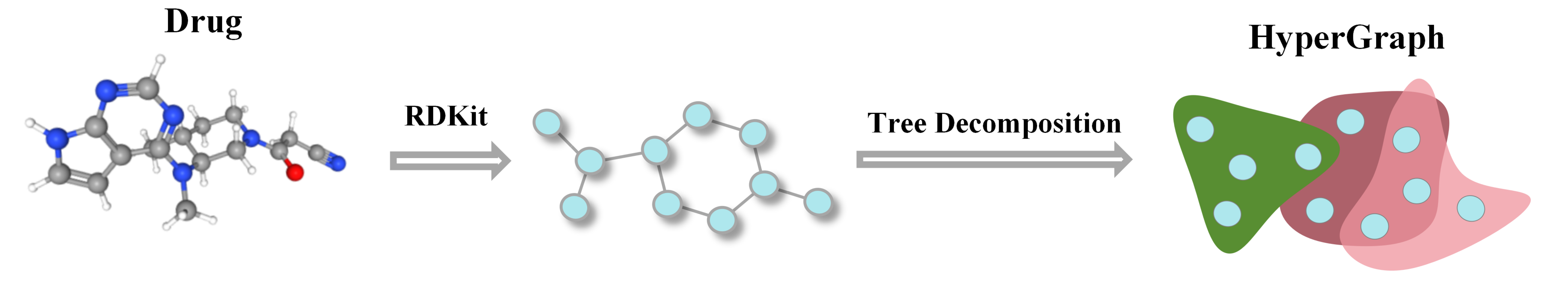}
    \caption{Schematic diagram of hypergraph construction process}
    \label{fig:hyper}
\end{figure}

There are various methods for constructing hypergraphs, including distance-based methods, attribute-based methods, and representation-based methods\cite{cosconati2010virtual}. Each method has its advantages and disadvantages, and the most suitable hypergraph construction method can only be selected for a specific task. To adapt to the Drug-Target Affinity (DTA) task, we introduce the tree decomposition algorithm\cite{jin2018junction} for hypergraph construction. The construction process is illustrated in Figure \ref{fig:hyper}.

Hypergraph neural networks are neural networks designed based on hypergraph structures. They aggregate node information using hyperedge convolution operations and leverage hyperedge connections to update node features. The HCAF-DTA model adopts the HypergraphConv model proposed in\cite{bai2021hypergraph} to extract features from the constructed hypergraphs. In the hyperedge matrix $H_c$, each hyperedge represents a substructure, and hypergraph neural networks can capture and express higher-order interactions between substructures. However, constructing hypergraphs alters the original semantics of molecular graphs, causing the features extracted by hypergraph neural networks to lack original semantic information. To compensate for this shortcoming, the model designs a hybrid model combining graph neural networks and hypergraph neural networks, serving as the drug encoder $f_c(\cdot)$. First, the model inputs the original semantic information of molecules obtained from SMILES sequences, i.e., the feature matrix $X_c$ and adjacency matrix $A_c$, into a three-layer GraphConv model and employs a gated skip-connection mechanism\cite{ryu2018deeply} for multi-scale feature extraction:

\begin{align}
X_c^{(l)} &= GNN_l(X_c^{(l-1)}, A_c), \nonumber \\
M_c &= \sigma\left(X_c^{(l)}W_{c, 1} + X_c^{(l-1)}W_{c, 2} + b_{z}\right) \label{skip1} \\
X_c^{(l)} &= M_c \odot X_c^{(l-1)} + \left(1 - M_c\right) \odot X_c^{(l)} \label{skip2}
\end{align}

Here, $GNN_l$ represents the $i$-th layer of GraphConv convolution, $W_{c,1}$ and $W_{c,2}$ are learnable weight matrices, and $b_c$ is a learnable bias term.

The original feature matrix $X_c$ is used as $X_c^{(0)}$. By propagating features through the above steps and retaining features from different layers using skip connections, the final feature matrix $X_c^{(3)}$ containing original semantic information is obtained. Subsequently, $X_c$ and $X_c^{(3)}$ are concatenated to form $X_h^{(0)}$, which, along with the hyperedge matrix $H_c$, is input into a three-layer HypergraphConv model for substructure feature extraction. Skip connections are also used to retain features at different scales:

\begin{align}
X_h^{(0)} &= X_c^{(3)} \parallel X_c, \nonumber \\
X_h^{(l)} &= \sigma(D_h^{-1}H_cWB_h^{-1}H_c^{T}X_h^{(l-1)}\Theta^{(l)}), \nonumber \\
X_h^{(l)} &= \text{Skip-Connection}(X_h^{(l)}, X_h^{(l-1)}), \nonumber
\end{align}

where $B_h$ is the degree matrix of hyperedges, $D_h$ is the degree matrix of nodes in the hypergraph, $W$ is the weight matrix of hyperedges, $\Theta^{(l)}$ is the learnable weight matrix of the $l$-th layer, and the Skip-Connection follows the same calculation method as Equations \ref{skip1} and \ref{skip2}.

Finally, the model concatenates the features $X^{(3)}$ containing original semantic information and the features $X_h^{(3)}$ containing substructure information to obtain the final molecular embedding feature matrix:

\begin{equation}
Z_c = X_c^{(3)} \parallel X_h^{(3)}, \nonumber
\end{equation}

where $\parallel$ denotes the concatenation operation.
\subsection{Spatial feature extraction of proteins}
Proteins are typically represented as 1D sequences composed of 25 distinct amino acids. However, relying solely on 1D sequence representation is insufficient for understanding the complex spatial conformations of proteins. Therefore, this study constructs protein graphs from protein sequences by leveraging the ESM-1b model\cite{rao2020transformer} to predict protein contact maps. The ESM-1b model is an unsupervised protein language modeling approach based on a Transformer architecture, which explores large-scale protein sequences and structures through pre-training. It can accurately and efficiently predict protein contact maps by directly inputting 1D protein sequences.

After obtaining the protein graph, the model constructs a three-layer Graph Neural Network (GNN) and employs a gated skip-connection mechanism for multi-scale feature extraction. For the given initial protein matrix $X^{(0)}_p$ and adjacency matrix $A_p$, feature extraction is performed as follows:

\begin{align}
X_p^{(l)} &= GNN_l(X_p^{(l)}, A_p), \nonumber \\
X_p^{(l)} &= \text{Skip-Connection}(X_p^{(l)}, X_p^{(l-1)}), \nonumber
\end{align}

where $GNN_l$ denotes the $i$-th layer of the GNN, and $\text{Skip-Connection}$ follows the same calculation method as Equations \ref{skip1} and \ref{skip2}. The output of the last layer is $X_p^{(3)}$. For the sake of subsequent discussion, this study denotes it as $Z_p$.
\subsection{Bi-Directional Multi-Cross Attention Fusion}
In the field of biology, the interaction between drugs and proteins is often metaphorically described as a lock-and-key relationship, where the protein acts as the lock to be unlocked, and the drug serves as the key that must precisely match to achieve binding. Molecular docking technology is based on this model\cite{cosconati2010virtual} and aims to simulate the interaction between drugs and proteins. However, many current end-to-end models merely concatenate the features of drugs and proteins linearly and then use a Multi-Layer Perceptron (MLP) to identify potential relationships between them. This approach fails to establish a deep connection between drugs and proteins and also ignores the complexity of actual interactions between atoms and amino acids.

\begin{figure}[htbp]
    \vspace{-10pt}
    \centering
    \includegraphics[width=1\linewidth]{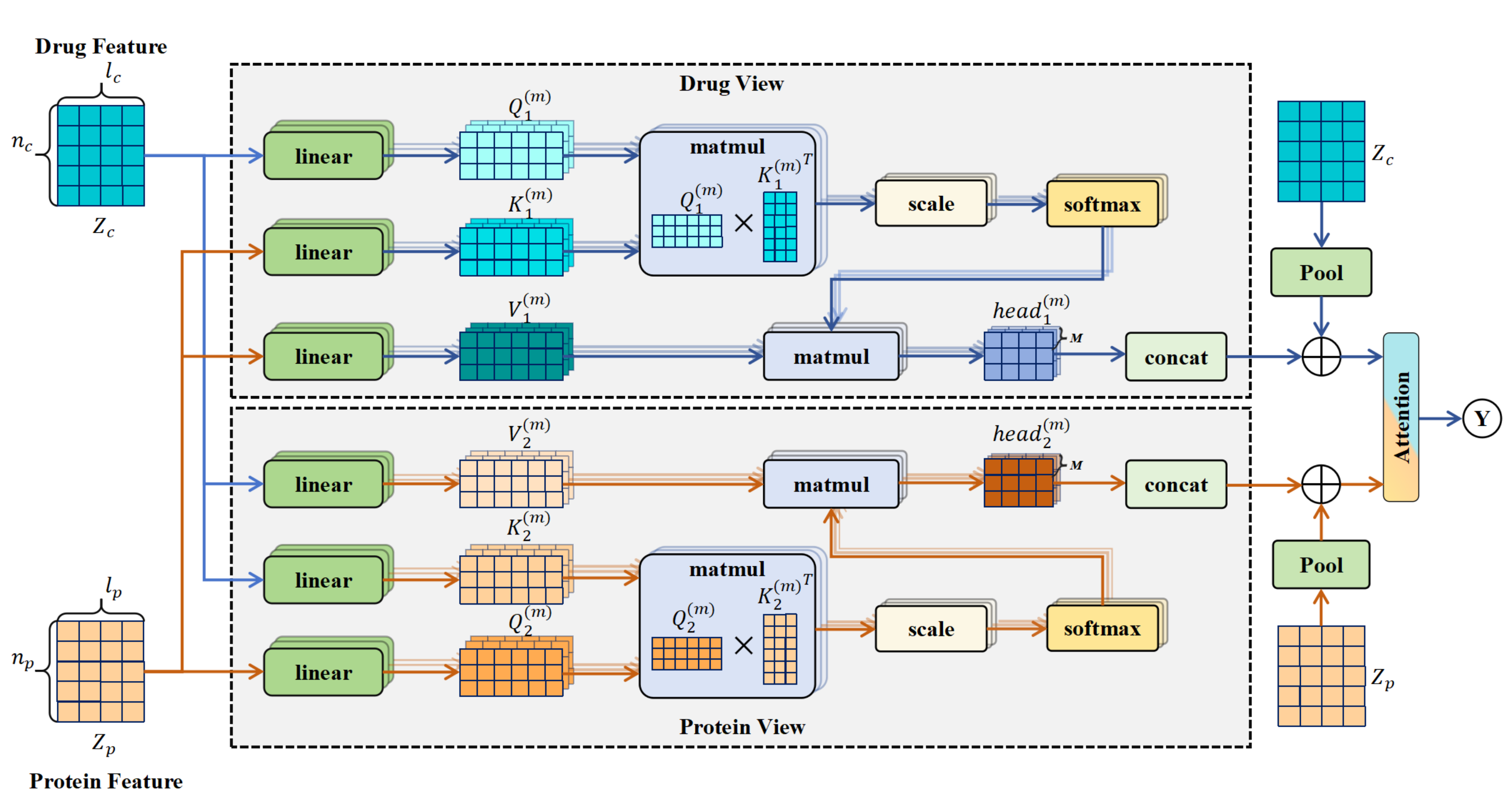}
    \caption{Bi-directional multi-head cross-attention module}
    \label{fig:cross_attention}
    \vspace{-10pt}
\end{figure}

The HCAF-DTA model designs a bidirectional multi-head cross-attention module, as shown in Figure \ref{fig:cross_attention}, which can fuse the features of each atom and amino acid and capture complex interaction relationships from the perspectives of atoms and amino acids, respectively.

From the perspective of atoms, the contribution of different residues in the protein to each atom in the drug is estimated. The drug embedding feature matrix $Z_c$ is used to generate the query matrix, and the protein embedding feature matrix $Z_p$ is used to generate the key and value matrices through linear mapping:
\begin{align}
Q_1^{(m)} = Z_c W_{Q_1}^{(m)}, \quad K_1^{(m)} = Z_p W_{K_1}^{(m)}, \quad V_1^{(m)} = Z_p W_{V_1}^{(m)}, \nonumber
\end{align}
where $m \in \{1, 2, \ldots, M\}$, $Z_c \in \mathbb{R}^{n_c \times l_c}$ and $Z_p \in \mathbb{R}^{n_p \times l_p}$ are extracted by the drug encoder $f_c(\cdot)$ and protein encoder $f_p(\cdot)$, respectively. $W_{Q_1}^{(m)} \in \mathbb{R}^{l_c \times l_q}$, $W_{K_1}^{(m)} \in \mathbb{R}^{l_p \times l_k}$, and $W_{V_1}^{(m)} \in \mathbb{R}^{l_p \times l_v}$ are weight matrices, where $l_q = \frac{l_c}{M}$, $l_k = l_v = \frac{l_p}{M}$. The attention mechanism is then used to calculate the contribution values by computing the dot product of all keys and queries, dividing each key by $\sqrt{l_q}$, and applying the softmax function to obtain the weights of the values:
\begin{align}
\text{head}_1^{(m)} = \text{softmax}\left(\frac{Q_1^{(m)} {K_1^{(m)}}^T}{\sqrt{l_q}}\right) V_1^{(m)}, \nonumber
\end{align}
Finally, all outputs are concatenated and a linear transformation is performed to obtain the output features:
\begin{align}
F_c=(\parallel_{m=1}^M head_1^{(m)})W_c, \nonumber
\end{align}

where $\parallel_{m=1}^M$ denotes the concatenation operation, and $W_c\in\mathbb{R}^{l_c \times l_c}$ is a weight matrix.

From the perspective of the protein, the contribution of different atoms in the drug to each residue in the protein is estimated. The protein embedding feature matrix $Z_p$ is used to generate the query matrix, and the drug embedding feature matrix $Z_c$ is used to generate the key and value matrices. Following the same process as above, the output feature $F_p \in \mathbb{R}^{n_p \times l_c}$ is obtained.

The extracted $S_c$ and $S_p$ contain complex interaction information between drugs and proteins. To enrich the representation information used for prediction, this study combines the features containing self-structural information of drugs and proteins with the above features:
\begin{align}
F_c = \lambda_c Z_c + (1 - \lambda_c) F_c, \quad F_p = \lambda_p Z_p + (1 - \lambda_p) F_p, \nonumber
\end{align}
where $\lambda_c$ and $\lambda_p$ are tunable parameters.

During the prediction phase, the model uses max-pooling to extract graph-level representations from $F_c$ and $F_p$, and employs attention mechanisms to calculate contribution scores $\alpha_c$ and $\alpha_p$. Finally, the contribution scores are dynamically assigned to the features and input into an MLP for drug-target affinity prediction:
\begin{align}
\tilde{y} = \text{MLP}\left(\text{maxpool}(F_c) \alpha_c \parallel \text{maxpool}(F_p) \alpha_p\right), \nonumber
\end{align}

During the model training phase, the Mean Squared Error (MSE) is used as the loss function:
\begin{equation}
\mathcal{L}_{\text{MSE}} = \frac{1}{n} \sum_{i=1}^n (\tilde{y}_i - y_i)^2,
\end{equation}
where $\tilde{y}$ denotes the predicted vector, $y$ denotes the actual output vector, and $n$ is the number of samples.
\section{Experiments and Discussion}
\subsection{Datasets}
In this study, we conducted a comprehensive performance evaluation of the HCAF-DTA model on two widely recognized and publicly available datasets: Davis\cite{davis2011comprehensive} and KIBA\cite{tang2014making}. The Davis dataset provides 30,056 samples of interactions between 68 drugs and 442 targets, with binding affinities measured as dissociation constants (KD) ranging from 0.016 to 10,000. Due to the wide numerical range, the values were converted to logarithmic scale using the formula shown in Equation \ref{equation4}. The KIBA dataset includes binding affinities for 2,116 drugs and 229 targets, measured as KIBA scores ranging from 0.0 to 17.2. In terms of the number of drug-target pairs with interactions, the KIBA dataset is approximately four times larger than the Davis dataset.

\begin{equation}
\mathrm{pK_d = -\log_{10}\left(\frac{K_d}{10^9}\right)} \label{equation4}
\end{equation}

To ensure a fair comparison of model performance, the dataset splitting method for Davis and KIBA followed the approach used in GraphDTA\cite{nguyen2021graphdta}. Table \ref{table1} provides more detailed information about these datasets.

\begin{table}[htbp]
    \centering
    \caption{Detailed information of the Davis and KIBA datasets}
    \label{table1}
    \begin{tabular}{ccccc}
        \toprule
DataSet      & Compounds & Proteins & Interactions \\ 
        \midrule
Davis        & 68        & 442      & 30056 \\
KIBA         & 2111      & 229      & 118254    \\
        \bottomrule
    \end{tabular}
\end{table}
\subsection{Hyperparameter Settings and Evaluation Metrics}
To compare the performance of our model with baseline models in regression tasks, we evaluated the model using three metrics: Mean Squared Error (MSE), $r_m^2$, and the Concordance Index (CI)\cite{gonen2005concordance}. The CI index measures the discriminative ability between different models on a scale from 0 to 1, with higher CI values indicating better model fit. The $r_m^2$ metric, proposed in DeepDTA, is used to assess the external prediction performance of the model. Table \ref{table2} defines the hyperparameters used in the HCAF-DTA model, where Graph Neural Network (GNN) layers utilize the GAT architecture\cite{velivckovic2017graph}. To ensure comparability, common hyperparameters for training deep learning models, such as learning rate and number of epochs, were set consistently with those in GraphDTA\cite{nguyen2021graphdta}.
\begin{table}[htbp]
    \centering
    \caption{Hyperparameters used in the HCAF-DTA model}
    \label{table2}
    \begin{tabular}{cc}
        \toprule
Hyper-paramet      & Settings \\ 
        \midrule
Learning rate        & 0.0005 \\
Batch size         & 512  \\
Optimizer         & Adam  \\
GAT layers(Drugs)  & 3  \\
GAT layers(Proteins)  & 4  \\
 HypergraphConv layers  & 2  \\
        \bottomrule
    \end{tabular}
\end{table}
\subsection{Comparison Experiments}
To comprehensively evaluate the performance of our proposed HCAF-DTA model, we compared it with the following models: traditional deep learning-based methods WideDTA and DeepDTA; graph-based methods GraphDTA, WGNN-DTA, DGraphDTA, and GSAML-DTA.
\begin{table*}[htbp]
\caption{Comparison of model prediction performance on the Davis and KIBA datasets \label{hyperCLresult}}
\tabcolsep=0pt
\begin{tabular*}{\textwidth}{@{\extracolsep{\fill}}lcccccc@{\extracolsep{\fill}}}
\toprule%
& \multicolumn{3}{@{}c@{}}{Davis} & \multicolumn{3}{@{}c@{}}{KIBA} \\
\cline{2-4}\cline{5-7}%
{\small Model} & {\small MSE} & {\small CI} & {\small $r_m^2$} & {\small MSE} & {\small CI} & {\small $r_m^2$} \\
\midrule
{\small DeepDTA}      & {\small 0.261(0.007)}          & {\small 0.878(0.004)}          & {\small 0.630(0.017)}          & {\small 0.194(0.008)}          & {\small 0.863(0.002)}          & {\small 0.673(0.019)}          \\
{\small WideDTA}      & {\small 0.262(0.009)}          & {\small 0.886(0.003)}          & {\small 0.633(0.011)}          & {\small 0.179(0.008)}          & {\small 0.875(0.001)}          & {\small 0.675(0.004)}          \\
{\small GraphDTA}     & {\small 0.229(0.005)}          & {\small 0.893(0.002)}          & {\small 0.685(0.016)}          & {\small 0.139(0.008)}          & {\small 0.889(0.001)}          & {\small 0.725(0.018)}          \\
{\small DGraphDTA}      & {\small 0.202}          & {\small 0.904}          & {\small 0.700}          & {\small 0.126}          & {\small 0.904}          & {\small 0.786}          \\
{\small WGNN-DTA(GCN)}      & {\small 0.208}          & {\small 0.900}          & {\small 0.692}          & {\small 0.144}          & {\small 0.885}          & {\small 0.781}          \\
{\small WGNN-DTA(GAT)}    & {\small 0.208}                 & {\small 0.903}          & {\small 0.691}          & {\small 0.130}                 & {\small 0.898}          & {\small 0.791}                  \\
{\small GSAML-DTA} & {\small 0.201(0.003)}          & {\small 0.896(0.001)}         & {\small 0.718(0.004)}  & {\small 0.132(0.002)}          & {\small 0.900(0.004)} & {\small 0.800(0.004)} \\
{\small HCAF-DTA} & {\small \textbf{0.198(0.002)}} & {\small \textbf{0.908(0.001)}} & {\small \textbf{0.728(0.004)}} & {\small \textbf{0.122(0.002)}} & {\small \textbf{0.907(0.003)}}          & \textbf{{\small 0.802(0.003)}}      \\
\bottomrule
\end{tabular*}
\end{table*}

The training setup involved repeating experiments with five different random seeds, calculating the mean and standard deviation of the results. The horizontal comparison of model performance is shown in Table \ref{hyperCLresult}. The table presents the experimental results of the HCAF-DTA model and all state-of-the-art models. It is evident that by extracting substructure features and drug-target interaction information, the HCAF-DTA model achieved the best performance across all datasets and evaluation metrics. The metrics on the Davis dataset were 0.198, 0.098, and 0.728, while those on the KIBA dataset were 0.122, 0.907, and 0.802.
Among the benchmark models, DGraphDTA, WGNN-DTA, and GSAML-DTA share the same input as the HCAF-DTA model, i.e., the original input consists of protein graphs and drug molecule graphs. The experimental results show that, under the same original feature input conditions, the HCAF-DTA model demonstrates superior predictive performance in the drug-target affinity (DTA) prediction task. Notably, the HCAF-DTA model showed significant performance improvements on the KIBA dataset, with a 4\% reduction in the MSE metric compared to the best-performing model, and improvements of 0.9\% and 0.6\% in the CI and $r_m^2$ metrics, respectively.
\subsection{Cold-start Experiments}
\begin{figure}[htbp]
\centering
\includegraphics[width=1\linewidth]{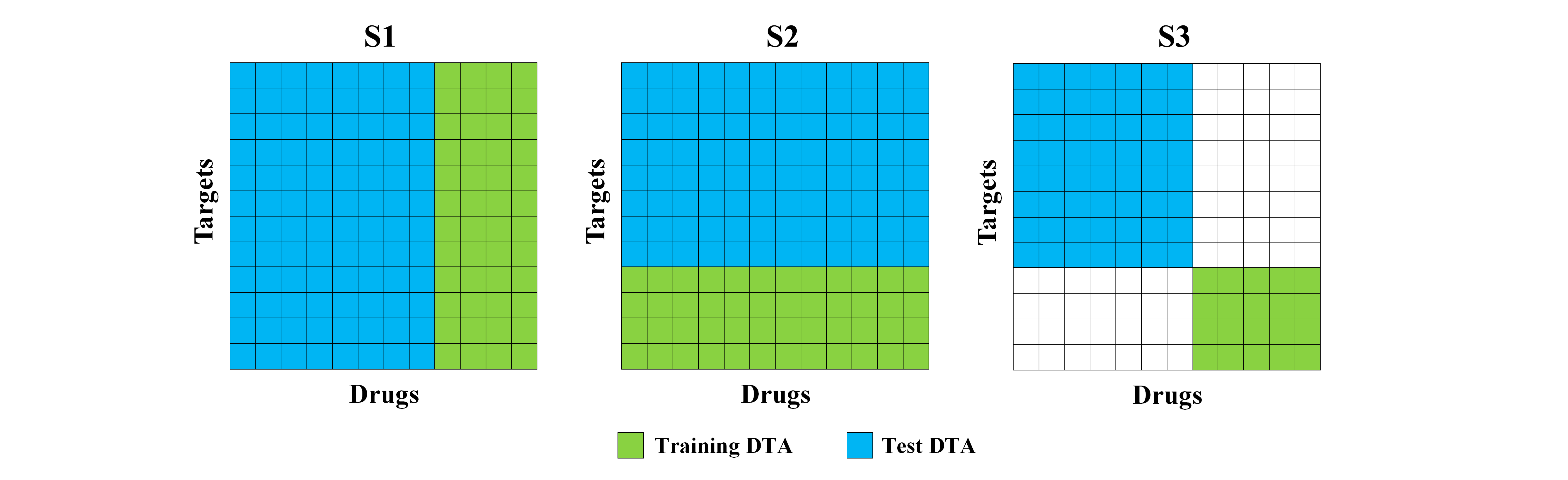}
\caption{Configuration comparison of three cold-start scenarios}
\label{fig:coldset}
\end{figure}

To comprehensively validate the model's generalizability and robustness, we additionally consider three experimental scenarios with specific configurations as illustrated in Figure \ref{fig:coldset}:
\begin{itemize}
\item \textbf{S1}: Random selection of column vectors from the drug-target affinity matrix for testing, with the remainder used for training.
\item \textbf{S2}: Random selection of row vectors from the drug-target affinity matrix for testing, with the remainder used for training.
\item \textbf{S3}: Testing on the intersection of row vectors from \textbf{S1} and column vectors from \textbf{S2}, where non-overlapping portions are excluded from both training and testing.
\end{itemize}

Under \textbf{S1} configuration, the model encounters known protein sequences but unknown drug sequences, simulating affinity target prediction for novel drugs. In this scenario, each unique drug sequence appears exclusively in either the training, validation, or test set. The \textbf{S2} setting presents known drug sequences with unknown protein sequences, replicating drug repurposing scenarios for novel targets, where each protein sequence remains exclusive to one data partition. \textbf{S3} represents the most challenging case where both drug and protein sequences are novel to the model, implemented through combinatorial partitioning that ensures no shared drugs or targets across training, validation, and test sets.

We evaluate these scenarios on the Davis dataset with 8:2 splits. For \textbf{S1}, 54/14 drugs are allocated to training/testing respectively. In \textbf{S2}, 354/88 proteins are divided accordingly. \textbf{S3} combines these splits, yielding training/test pairs of (354 proteins × 54 drugs)/(88 proteins × 14 drugs). Five-fold cross-validation results with numerical distribution of drug-target pairs are presented in Table \ref{tablecold}.

\begin{table}[htbp]
\caption{Dataset partitioning results of five-fold cross-validation under three scenarios.\label{tablecold}}
\centering
\begin{tabular}{ccccccc}
\toprule%
\textbf{Split}    & \textbf{Setting} & \textbf{Fold1} & \textbf{Fold2} & \textbf{Fold3} & \textbf{Fold4} & \textbf{Fold5} \\ \hline
Training & S1      & 23868 & 23868 & 23868 & 24310 & 24310 \\
         & S2      & 23936 & 23528 & 24276 & 23936 & 24548 \\
         & S3      & 19008 & 18684 & 19278 & 19360 & 19855 \\
Test     & S1      & 6188  & 6188  & 6188  & 5746  & 5746  \\
         & S2      & 6120  & 6528  & 5780  & 6120  & 5508  \\
         & S3      & 1260  & 1344  & 1190  & 1170  & 1053  \\ \bottomrule
\end{tabular}
\end{table}

\begin{figure}[htbp]
\centering
\includegraphics[width=0.8\linewidth]{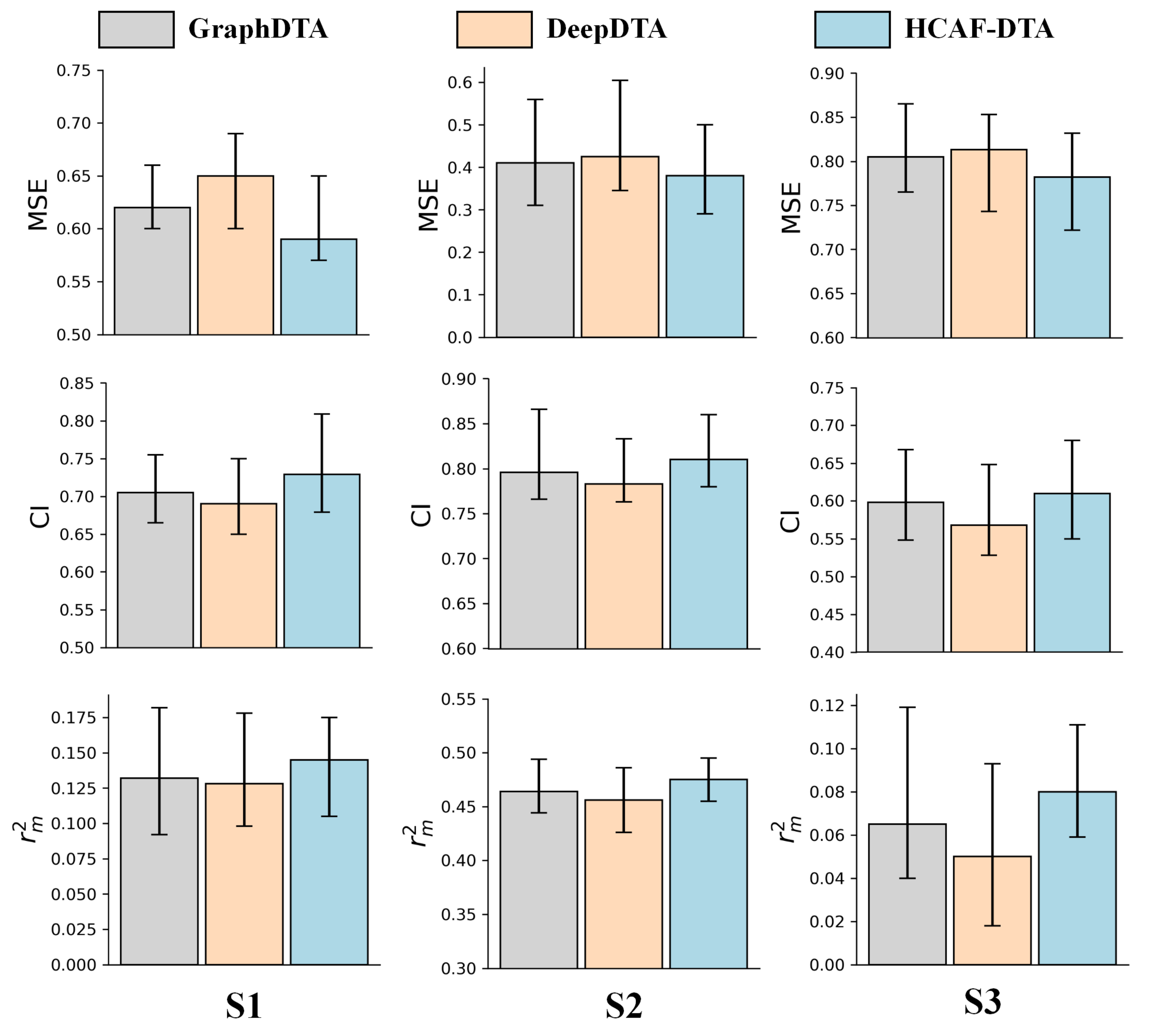}
\caption{Performance comparison of cold-start experiments}
\label{fig:coldtest}
\end{figure}

Figure \ref{fig:coldtest} presents horizontal performance comparisons of DeepDTA, GraphDTA, and HCAF-DTA across scenarios. Column bars represent mean values under five-fold cross-validation. The results demonstrate that: 1) All models perform worst in \textbf{S3} due to complete novelty of both drug and target sequences; 2) Superior performance in \textbf{S2} versus \textbf{S1} reflects richer protein feature learning from the protein-abundant Davis dataset; 3) HCAF-DTA consistently outperforms baselines across all scenarios, confirming its enhanced generalizability and robustness.

\subsection{Ablation Study}
To capture which amino acids and atoms play critical roles in drug-target interaction processes, the HCAF-DTA model incorporates a bidirectional multi-head cross-attention mechanism. Additionally, the model introduces a hypergraph neural network to capture drug substructure features, thereby enriching molecular representations. To investigate key factors influencing model prediction performance, this study conducted comprehensive ablation experiments on the KIBA dataset by removing or replacing specific modules to generate variants of HCAF-DTA. For clarity, the bidirectional multi-head cross-attention module is denoted as Fusion, while the hypergraph neural network module is termed HyperDrug. When the Fusion module was removed, it was replaced with a simple concatenation operation.
\begin{table*}[htbp]
\caption{Performance Comparison of Ablation Experiments for the HCAF-DTA Model\label{tablehyperCLfu}}
\tabcolsep=0pt
\begin{tabular*}{\textwidth}{@{\extracolsep{\fill}}lccccc@{\extracolsep{\fill}}}
\toprule%
& \multicolumn{2}{@{}c@{}}{Module} & \multicolumn{3}{@{}c@{}}{Result} \\
\cline{2-3}\cline{4-6}%
Model & Fusion & HyperDrug & MSE & CI & $r_m^2$ \\
\midrule
Model1      & \checkmark                             & ×                          &  0.124(0.003)          & 0.903(0.004)          & 0.801(0.004)           \\
Model2      & ×                             & \checkmark                                      & 0.125(0.003)          & 0.903(0.004)          & 0.802(0.002)          \\
Model3      & ×                             & ×                                             & 0.124(0.002)          & 0.905(0.003)          & 0.801(0.003) \\
HCAF-DTA & \checkmark                             & \checkmark                          & \textbf{0.121(0.002)} & \textbf{0.908(0.003)} & \textbf{0.805(0.003)} \\
\bottomrule
\end{tabular*}
\end{table*}

The ablation results are presented in Table \ref{tablehyperCLfu}. The table clearly demonstrates that the HCAF-DTA model exhibits superior performance compared to other variants. The worst performance occurred when both modules were removed (see Model 3). Furthermore, this study observed performance degradation when either the Fusion module (Model 2) or the HyperDrug module (Model 1) was ablated. This indicates that both modules enhance drug feature representations—either by enriching molecular characteristics or by extracting fused feature information—thereby improving the model's predictive capability.
\section{Conclusion}
To address the limitations of current drug-target affinity prediction methods, including restricted extraction of higher-order molecular features, missing protein spatial information, and insufficient modeling of intermolecular interactions, this chapter proposes HCAF-DTA, a novel model based on hypergraph and bidirectional multi-head attention contrastive learning. The model aims to enhance feature representation and generalization performance in cold-start scenarios.

HCAF-DTA extracts multi-scale features by constructing hypergraphs for drug molecules and weighted graphs for proteins. For drug representation, hypergraphs built via tree decomposition capture substructure features using hyperedges, and skip connections integrate information from both hypergraphs and simple graphs. For protein representation, weighted graphs are constructed from residue contact maps predicted by the ESM model, and spatial structural features are extracted using multi-layer graph neural networks (GNNs). During prediction, the model uses a bidirectional multi-head cross-attention mechanism to establish fine-grained feature associations from both atomic and amino acid perspectives, effectively simulating complex interactions during molecular docking.

The experiments show that HCAF-DTA outperforms the comparison model on the Davis and KIBA datasets. In addition, the results of three cold-start experiments show that the model is highly generalisable and robust, which helps to advance the study of drug structure optimisation.
\bibliography{reference.bib}

\end{document}